\documentclass[showpacs,pra,twocolumn]{revtex4}
\usepackage{dcolumn}
\usepackage{mathrsfs}
\usepackage{amsmath}
\usepackage{graphicx}
\usepackage{bm}
\usepackage{epstopdf}
\usepackage{float}
\usepackage{hyperref}
\usepackage{graphicx,subfigure,xcolor}
\usepackage{braket}

\newcommand{\bea}{\begin{eqnarray}}
\newcommand{\eea}{\end{eqnarray}}
\newcommand{\beq}{\begin{equation}}
\newcommand{\eeq}{\end{equation}}
\newcommand{\nn}{\nonumber}

\def\/{\over}

\def\w{\omega}
\def\w0{\omega_{0}}

\def\bk{{\bf k}}

\def\br{{\mathbf r}}

\begin{document}

\title{Vacuum fluctuations and radiation reaction contributions to the resonance dipole-dipole interaction between two atoms near a reflecting boundary}
\author{Wenting Zhou$^{1,2,3}$\footnote{zhouwenting@nbu.edu.cn}}

\author{Lucia Rizzuto$^{3,4}$\footnote{lucia.rizzuto@unipa.it}}

\author{Roberto Passante$^{3,4}$\footnote{roberto.passante@unipa.it}}

\affiliation{$^{1}$Center for Nonlinear Science and Department of Physics, Ningbo
University, Ningbo, Zhejiang 315211, China}
\affiliation{$^2$China Key Laboratory of Low Dimensional Quantum Structures and Quantum Control of Ministry of Education, Hunan Normal University, Changsha,
Hunan 410081, China}
\affiliation{$^3$Dipartimento di Fisica e Chimica, Universit\`{a} degli Studi di Palermo, Via Archirafi 36, I-90123 Palermo, Italy}
\affiliation{$^4$INFN, Laboratori Nazionali del Sud, I-95123 Catania, Italy}

\begin{abstract}
We investigate the resonance dipole-dipole interaction energy between two identical atoms, one in the ground state and the other in the excited state, interacting with the electromagnetic field in the presence of a perfectly reflecting plane boundary. The atoms are prepared in a correlated (symmetric or anti-symmetric) Bell-type state. Following a procedure due to Dalibard {\em et. al.} [J. Dalibard {\it et. al.}, J. Phys. (Paris) {\bf 43}, 1617 (1982); {\bf 45}, 637 (1984)], we separate the contributions of vacuum fluctuations and radiation reaction (source) field to the resonance interaction energy between the two atoms, and show that only the source field contributes to the interatomic interaction, while vacuum field fluctuations do not.
By considering specific geometric configurations of the two-atom-system with respect to the mirror and specific choices of dipole orientations, we show that the presence of the mirror significantly affects the resonance interaction energy and that different features appear with respect to the case of atoms in free space, for example a change in the spatial dependence of the interaction.
Our findings also suggest that the presence of a boundary can be exploited to tailor and control the resonance interaction between two atoms, as well as the related energy transfer process. The possibility of observing these phenomena is also discussed.
\end{abstract}
\maketitle

\section{Introduction}
\label{Sec.1}

Radiative properties and radiative corrections of atoms, and quantum emitters in general, interacting with the quantized electromagnetic field have been extensively investigated in the literature, and it is now well understood that these processes can be controlled and tailored through the environment. In fact, the structure of field modes and of the photonic density of states can significantly change in a cavity or in a structured environment, thus affecting radiative properties of any quantum emitter inside it \cite{Purcell69, HK89,MK73,Yablonovitch87,JQ94}. The influence of boundary conditions on atomic radiative processes was first revealed in the seminal work of Purcell \cite{Purcell69}, who showed that the spontaneous emission rate of an atom can be enhanced or inhibited through the environment. This remarkable finding stimulated many subsequent theoretical and experimental investigations yielding the important conclusion that light-matter interactions can be successfully manipulated by a suitable choice of the environment. Changes in the lifetimes and radiative shifts of excited atoms have been extensively investigated and experimentally verified \cite{HK89} and it is now well known that the spontaneous emission of an excited atom or a quantum dot can be enhanced or suppressed if the atom is embedded in a nanostructured medium, for example a waveguide or a photonic crystal  \cite{LNNB00, AKP04, NFA07}.

Besides  affecting spontaneous emission, the environment can also modify radiation-mediated interactions between atoms, such as dispersion interactions between atoms (van der Waals and Casimir-Polder interactions) or between an atom and a macroscopic object  \cite{CP48, Milonni93, CPP95, SPR06,MK69,CT98, HIHSPPS09,HS15,PPR07}, or resonance interaction energy and energy transfer between two atoms \cite{Salam10,Forster65,JA00,ADG98}. For example, it has been shown that the resonance interaction energy between two identical entangled atoms can be enhanced or suppressed when the atoms are embedded in a nanostructured environment  with a photonic bandgap \cite{IFTPRP14,NPR17}. The effect of reflecting boundaries on the radiative properties of entangled atoms has also been investigated recently in the literature \cite{ADMS16}.

Very recently, the possibility to control radiative processes through a dynamical modulation of the environment has been explored too. For example, it has been shown that the spontaneous emission rate of an atom inside a photonic crystal, as well as  the spectrum of the emitted radiation, can be significantly modified if the environment is adiabatically modulated in time  \cite{CRP17,GHDPZ10}.
Dynamic control of the F\"{o}rster energy transfer in a photonic environment also  has been discussed recently \cite{SKKM14}.

In this paper we investigate the resonance interatomic interaction between two identical atoms, one in an excited state and the other in the ground state, prepared in a correlated Bell-type state, and interacting with the electromagnetic field in the vacuum state and in the presence of a perfectly reflecting plate.
Our aim is to discuss the effect of the presence of the reflecting plate on the resonance interaction energy, and whether it can be enhanced or suppressed through a suitable choice of the geometric configuration and dipole orientation of the two-atom-system  with respect to the mirror.
We also consider the separation of the interaction energy in terms of its vacuum fluctuations and radiation reaction contributions.

As it is known, radiative corrections of atoms embedded in a confined space, similarly to the free-space case, can be physically described in terms of both vacuum field fluctuations and the radiation reaction field. In the first case, the cavity effects are described as due to modifications of the structure of vacuum field fluctuations inside the cavity. In the second case, the same effects are interpreted as the interaction of the atom with the field emitted by itself, or by the other atom, and reflected by the cavity walls.
In the present work, we adopt something in between; specifically, following a procedure introduced by Dalibard {\em et. al} \cite{DDC82,DDC84}, we separate at the second order in perturbation theory the contributions of vacuum field fluctuations and the radiation reaction field to the total energy shift of the two-atom system near the reflecting plate.
Our approach will also permit us to deepen our understanding of the physical properties of the resonance interaction energy and the effects induced by the boundary.

Generally speaking, resonance interactions between identical atoms take place when one or both atoms are in an excited state, and an exchange of real photons between the atoms is involved. If the two-atom system is prepared in a factorized state (for example $\vert g_A\rangle\vert e_B\rangle$, atom $A$ in the ground state and atom $B$ excited), the interaction is a
fourth-order effect in the atom-field coupling, and it is the well-known Casimir-Polder interaction between excited atoms \cite{CP48}. This interaction is of very long-range range, scaling as $R^{-2}$ in the far-zone limit $R \gg \lambda$ ($\lambda$ being the relevant wavelength associated with the atomic transition frequency and $R$ the interatomic distance).
Recently, this interaction has been extensively investigated in the literature, also in connection with some controversial results concerning the presence of spatially oscillating terms and the consequent change of the resonance force from attractive to repulsive \cite{RPP04, Berman15, DGL15, BPRB16, MR15}. A different phenomenon occurs when two identical atoms are prepared in a correlated symmetric or antisymmetric state, with one atom in an excited state and the other in the ground-state, so that the excitation is delocalized and the atomic dipoles are correlated; in this case, the resonance interaction is a second-order effect, and it scales as $1/R$ in the far-zone limit \cite{CT98,Salam10}. Resonance interactions are thus stronger effects than dispersion interactions, even if they require the preparation of a correlated state of the two atoms. Together with the F$\ddot{o}$rster energy transfer \cite{Forster65}, the resonance interaction is relevant for many important processes in biophysics, such as photosynthesis \cite{JA00, PP13, RLBBLB14}, as well as in the formation of cold molecules  \cite{JJLPTW96, DDDLMP01}, cold collision processes \cite{FG06}, quantum storage and transmission \cite{KK06} and laser cooling \cite{Philips98}.

The resonance interaction between two uncorrelated atoms at  finite temperature \cite{BLMN03}, or in the presence of a dielectric medium \cite{BN04,YZY16,WPA15} or a mirror \cite{WPGA17} has been  recently investigated, as well as the effects of a nanostructured material such as microcavities \cite{KKY96}, nanofibers \cite{KGNH05,WPA16} and waveguides \cite{SK13}.
Also, the effect of non-inertial motion of the atoms on the resonance interaction, as well as on the Casimir-Polder interaction has been studied \cite{RLMNSZP16, ZPR16,NP13}.

As mentioned, in this paper we investigate the resonance interaction between two identical atoms prepared in a symmetric or antisymmetric correlated state, and located in proximity of an infinite perfectly reflecting plate.
We consider two different geometric configurations for the two-atom system, specifically two atoms in a perpendicular or parallel direction relative to the reflecting plate, and separate the contributions of vacuum fluctuations and source fields to the interaction energy. We show that the resonance interaction is exclusively due to the radiation reaction contribution, while the vacuum fluctuations contribution only gives the Lamb shift of the individual atoms. We compare our findings with the case of atoms in free space and show that specific features appear due to the presence of the boundary: a qualitative change of the interaction energy and a change of its strength. We discuss these results for different choices of the orientation of the atomic dipoles with respect to the wall.

The paper is organized as follows. In Sec. \ref{Sec.2} we introduce our model and the general approach used to evaluate the resonance interatomic interaction in terms of vacuum fluctuations and source fields.
In Sec. \ref{Sec.3} we evaluate the resonance interaction between two atoms for two different configurations: atoms perpendicular and parallel to the reflecting plate. Section \ref{Sec.4} is devoted to a summary and our conclusions.
Throughout the paper we adopt units such that $\hbar=c=1$.

\section{Vacuum fluctuations and radiation reaction contributions to the resonance interaction}
\label{Sec.2}

Let us consider two identical atoms $A$ and $B$ interacting with the electromagnetic field in the vacuum state and in the presence of a perfectly reflecting plate. We assume that the mirror is located at $z=0$. The atoms are modeled as two-level systems with eigenstates $\vert g\rangle$ and $\vert e\rangle$, corresponding to the energies $-\omega_0/2$ and $\omega_0/2$ respectively.
In this section we illustrate the method we use to separate vacuum fluctuations and radiation reaction contributions, introducing relevant notations and the expressions of atomic and field susceptibilities and symmetric correlation functions.
We work in the Heisenberg picture, and thus our operators have a time dependence. Also, for the sake of generality, in this Section we work in a more general context, including the possibility of atoms moving with the same proper acceleration and sharing the same proper time. In Sec. \ref{Sec.3}, we will specialize the expressions obtained in this section  to the specific case of two atoms at rest.

In the case of two atoms in a flat background spacetime and having the same proper time $\tau$ (as in the case we consider in this paper), in the Heisenberg picture, the Hamiltonian in the multipolar coupling scheme and within the dipole approximation, in the comoving (locally inertial) frame is given by\cite{CT98, CPP95,RLMNSZP16,MNP14}
\begin{eqnarray}
\label{eq:1}
H(\tau)&=&\omega_0\sigma^A_3(\tau)+\omega_0\sigma^B_3(\tau)\nonumber\\
&+&\sum_{\mathbf{k}\lambda}\omega_{k}a^{\dag}_{\mathbf{k}\lambda}(t(\tau))a_{\mathbf{k}\lambda}(t(\tau)){dt\/d\tau}\nonumber\\
&-&{\boldsymbol\mu_{A}(\tau)}\cdot\mathbf{E}(x_A(\tau))-\boldsymbol\mu_{B}(\tau)\cdot\mathbf{E}(x_B(\tau))\;,
\end{eqnarray}
with $\sigma_3={1\/2}(|e\rangle\langle e|-|g\rangle\langle g|)$, $\mathbf{k}$ and $\lambda$ the wave vector and polarization of the electromagnetic field respectively, $\boldsymbol\mu(\tau)=e\mathbf{r}$ the atomic dipole moment operator, and $\mathbf{E}(x(\tau))$ the electric field operator,

\begin{equation}
\label{eq:1a}
{\bf E}({x(\tau)})=-i\sum_{\bk \lambda}\sqrt{\frac{2\pi k}{V}}{\bf f}_{\bk\lambda}(\br)(a_{\bk\lambda}(t(\tau)) - a^{\dagger}_{\bk\lambda}(t(\tau)))\, ,
\end{equation}
where ${\bf f}_{\bk\lambda}(\br)$ are the mode functions satisfying the boundary conditions on the reflecting plate.
Here $x_A(\tau)$ and $x_B(\tau)$ are the trajectories of atoms $A$ and $B$ respectively, $\tau$ being the proper time.
We assume that the possible motion of the atoms is maintained by some external action on them. Although the proper time of an atom depends on its motion and the proper times of two moving atoms are in general different, the specific case considered here (same proper time) is not uncommon. For example, in the case of two atoms with the same uniform proper acceleration perpendicular to their (constant) separation, the proper time of the two atoms in the locally inertial comoving system is the same. Thus our Hamiltonian (\ref{eq:1}) is valid also in the more general case of atoms moving with the same proper acceleration and a constant separation and not only for the case of two atoms at rest in the laboratory system that we  investigate in this paper.
In the specific case of two atoms at rest in the laboratory system considered later on in this paper, the factor ${{dt}\/{d\tau}}$ reduces to unity. We wish to stress that in this section we are keeping this factor for the sake of generality, in order to obtain expressions valid also for more general and relevant cases, for example the mentioned case of atoms moving with uniform acceleration and constant separation.

As mentioned, in order to investigate  the resonance interaction energy we exploit a procedure proposed by Dalibard, {\it et. al.} \cite{DDC82},  which consists in separating the vacuum fluctuations and radiation reaction contributions to the energy shift of the two-atom system.
This approach has already been used to investigate the radiative properties of an atom in front of a reflecting plate \cite{MJH90}, or  atoms moving with uniform acceleration in vacuum  \cite{AM94, Passante98, Rizzuto07, ZY12, RS11, RS09, ZY10}, and very recently it has been generalized to the case of two atoms, to evaluate the resonance and dispersion interactions between two atoms in non inertial motion in the free space \cite{MNP14, RLMNSZP16, ZPR16, ZY17}.

We first solve the Heisenberg equations of motion for a generic atomic observable (pertaining to atom $A$ or $B$) at the second-order in the coupling constant, and separate the solutions in the vacuum fluctuation term (related to the free-field operators)  and  radiation reaction term (related to the presence of the field source). This leads to an effective Hamiltonian that governs the time evolution of the atomic observable
\begin{eqnarray}
\label{eq:1b}
(H^{eff}_{A})&=&(H^{eff}_{A})_{vf}+(H^{eff}_{A})_{rr}
\end{eqnarray}
where
\begin{eqnarray}
\label{eq:2}
(H^{eff}_{A})_{vf}&=&-{i\/2}\int^{\tau}_{\tau_0}d\tau'C^F_{ij}(x_A(\tau),x_A(\tau'))\nn\\
&&\times[{\mu}^{A}_i(\tau),{\mu}^{A}_j(\tau')]\;
\end{eqnarray}
is related to vacuum fluctuations, while
\begin{widetext}
\begin{eqnarray}
\label{eq:3}
(H^{eff}_{A})_{sr}=-{i\/2}\int^{\tau}_{\tau_0}d\tau'\chi^F_{ij}(x_A(\tau),x_A(\tau'))\{{\mu}^{A}_i(\tau),{\mu}^{A}_j(\tau')\}
-{i\/2}\int^{\tau}_{\tau_0}d\tau'\chi^F_{ij}(x_A(\tau),x_B(\tau'))\{{\mu}^{A}_i(\tau),{\mu}^{B}_j(\tau')\}\;.
\end{eqnarray}
\end{widetext}
is the contribution of radiation reaction field or source field. Similar expressions are obtained for atom $B$.
Here $[ , ]$ and $\{ , \}$ denote the commutator and anticommutator respectively, and we have introduced the field statistical functions $C^F_{ij}$ (symmetric correlation function) and $\chi^F_{ij}$  (linear susceptibility)
\begin{eqnarray}
\label{eq:4}
C^F_{ij}(x(\tau),x(\tau'))&=&{1\/2}\langle0|\{\mathrm{E}_{i}(x(\tau)),\mathrm{E}_{j}(x(\tau'))\}|0\rangle\;,\\
\label{eq:4a}
\chi^F_{ij}(x(\tau),x(\tau'))&=&{1\/2}\langle0|[\mathrm{E}_{i}(x(\tau)),\mathrm{E}_{j}(x(\tau'))]|0\rangle\;.
\end{eqnarray}

The vacuum fluctuations and radiation reaction contributions to the energy shift of the system are then obtained by evaluating
the average value of the effective Hamiltonians (\ref{eq:2}) and (\ref{eq:3}) on the atomic state of our two-atom system.

The procedure outlined above is very general. We now specialize our considerations to the case of two identical atoms prepared in a correlated Bell-type state. Specifically, we consider two atoms prepared in one of the two correlated states $|\psi_{+}\rangle$ or $|\psi_{-}\rangle$
(symmetric or antisymmetric)
\begin{eqnarray}
\label{eq:5}
|\psi_{\pm}\rangle={1\/\sqrt{2}}(|g_A,e_B\rangle\pm|e_A,g_B\rangle)\; .
\end{eqnarray}
In the Dicke model \cite{Dicke} the symmetric state is the  superradiant state, while the antisymmetric state is the subradiant state.
The total energy shift of the two-atom system is then obtained by evaluating the average value of the effective Hamiltonians (\ref{eq:2}) and  (\ref{eq:3}) on one of the two states (\ref{eq:5}),
\begin{eqnarray}
\label{eq:6}
(\delta E_A)_{vf}=-i \int^{\tau}_{\tau_0}d\tau'C^F_{ij}(x_A(\tau),x_A(\tau'))\chi^A_{ij}(\tau,\tau')\, ,
\end{eqnarray}
\begin{eqnarray}
\label{eq:7}
&&(\delta E_A)_{sr}=-i\int^{\tau}_{\tau_0}d\tau'\chi^F_{ij}(x_A(\tau),x_A(\tau'))C^A_{ij}(\tau,\tau')\label{sr-A}\nn\\
&&\quad\quad-i\int^{\tau}_{\tau_0}d\tau'\chi^F_{ij}(x_A(\tau),x_B(\tau'))C^{A,B}_{ij}(\tau,\tau')\, .
\end{eqnarray}
where $\chi^{A,B}_{ij}$ and $C^{A,B}_{ij}$ are  the antisymmetric and symmetric statistical functions of the atoms, respectively, defined as
\begin{eqnarray}
\label{eq:8}
\chi^{AB}_{ij}(\tau,\tau')&=&{1\/2}\langle\psi_{\pm}|[{\mu}^{A}_{i}(\tau),{\mu}^{B}_{j}(\tau')]|\psi_{\pm}\rangle\;,\\
\label{eq:8a}
C^{A,B}_{ij}(\tau,\tau')&=&{1\/2}\langle\psi_{\pm}|\{{\mu}^{A}_{i}(\tau),{\mu}^{B}_{j}(\tau')\}|\psi_{\pm}\rangle\;.
\end{eqnarray}

As discussed in \cite{RLMNSZP16} the vacuum fluctuations contribution (see Eq. (\ref{eq:6})) does not depend on the interatomic distance. It contributes only to the Lamb shift of each atom, as if the other atom were absent; therefore it is not relevant in the calculation of the resonance interaction between atoms. Similar considerations apply to the first term in (\ref{eq:7}), describing the contribution of the radiation reaction field to the Lamb-shift of each atom ($A$ or $B$). The only term relevant for the resonance interaction energy is that depending on the presence of both atoms in (\ref{eq:7}). Therefore, the resonance interaction between two correlated atoms is exclusively related to the radiation reaction term (at the second order in the coupling) \cite{RLMNSZP16}. This is expected from a physical point of view, because the correlated states (\ref{eq:5}) have non-vanishing dipole-dipole correlations, and
this is at variance with dispersion interactions (van der Waals and Casimir-Polder interactions),
where instantaneous dipole moments are induced in both atoms by the spatially correlated vacuum field fluctuations \cite{SPR06, PS07}.
Finally, for the distance-dependent terms, we obtain
\begin{eqnarray}
\label{eq:9}
\delta E&=&-i\int^{\tau}_{\tau_0}d\tau'\chi^F_{ij}(x_A(\tau),x_B(\tau'))C^{A,B}_{ij}(\tau,\tau')\nn\\&&+(A\rightleftharpoons B)\;.
\end{eqnarray}

This expression is very general and can be applied to evaluate the resonance interaction between two atoms in different situations, for example two uniformly accelerated atoms in vacuum \cite{RLMNSZP16} or in the presence of an external environment, provided the appropriate expression of the field susceptibility is evaluated.

\section{Resonance interaction between two atoms at rest near a perfectly reflecting boundary}
\label{Sec.3}

We now apply the procedure introduced in the preceding section to study the resonance interaction between two atoms placed near a perfectly conducting plate.  We assume the mirror to be in the ``$xOy$'' plane, with the $z$ axis perpendicular to the plate. The two atoms are located in the half space $z>0$. In order to simplify our discussion, we consider two different configurations of the system, specifically  two atoms aligned along the $z $axis, perpendicular to the boundary, and two atoms located along a direction parallel to the plate. This will permit us to simplify the calculation and to highlight some relevant effects of the plate on the resonant interaction energy between the two atoms.

\subsection{Two atoms aligned perpendicularly to the plate}
\begin{figure}[H]
\centering
\includegraphics[scale=0.45]{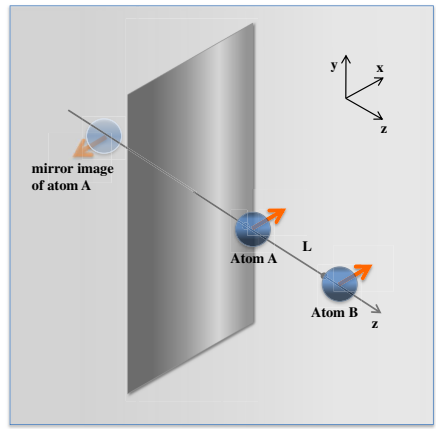}
\caption{Pictorial representation of the physical system of two atoms aligned along the $z$ axis, perpendicular to the plate.}\label{Fig1}
\end{figure}
Let us first consider both atoms fixed and located in the $z$ direction,  perpendicular to the surface, as shown in Fig. \ref{Fig1}.

In order to obtain the distance-dependent energy shift of the two-atom system, we first evaluate the
two-point correlation function of the electric field operator on the atomic trajectories $x_A(\tau_A)=(t_A,x_A,y_A,z_A)$ and $x_B(\tau_B)=(t_B,x_B,y_B,z_B)$  \cite{RS09, ZY10}
\begin{eqnarray}
\label{eq:10}
g_{ij}(x_A(\tau_A),x_B(\tau_B))=\langle0|\mathrm{E}_{i}(x_A(\tau_A))\mathrm{E}_{j}(x_B(\tau_B))\vert0\rangle\, .
\end{eqnarray}
This function can be written as sum of a free-space term, $g^{(0)}(x_a,x_b)$, and a boundary-dependent term $g^{(b)}(x_a,x_b)$
\begin{eqnarray}
\label{eq:11}
&\ & g_{ij}(x_A(\tau_A),x_B(\tau_B)) = g^{(0)}_{ij}(x_A(\tau_A),x_B(\tau_B)) \nonumber \\
&+& g^{(b)}_{ij}(x_A(\tau_A),x_B(\tau_B))
\end{eqnarray}
with
\begin{widetext}
\begin{eqnarray}
\label{eq:12}
g^{(0)}_{ij}(x_A(\tau_A),x_B(\tau_B))=-{1\/4\pi^2}(\delta_{ij}\partial_0\partial_{0'}-\partial_i\partial_{j'})
{1\/{(t_A-t_B-i\epsilon)^2-(x_A-x_B)^2-(y_A-y_B)^2-(z_A-z_B)^2}}\;,
\end{eqnarray}
\begin{eqnarray}
\label{eq:13}
g^{(b)}_{ij}(x_A(\tau_A),x_B(\tau_B))={1\/4\pi^2}[(\delta_{ij}-2n_in_j)\partial_0\partial_{0'}-\partial_i\partial_{j'}]
{1\/{(t_A-t_B-i\epsilon)^2-(x_A-x_B)^2-(y_A-y_B)^2-(z_A+z_B)^2}}\;,
\end{eqnarray}
\end{widetext}

where $\mathbf{n}=(0,0,1)$, $\partial_{0}=\frac{\partial}{\partial t}$, $\partial_{i}=\frac{\partial}{\partial x_i}$, and the prime refers to derivatives with respect to the coordinates $(t_B,x_B,y_B,z_B)$.  The field linear susceptibility is then obtained from Eqs. (\ref{eq:12}) and (\ref{eq:13}).

For two atoms at rest aligned perpendicularly to the boundary and separated by a distance $L$, as shown in Fig. \ref{Fig1}, the atomic trajectories $x_A(\tau_A)$ and $x_B(\tau_B)$ are
\begin{eqnarray}
\label{eq:14}
&&t_A=\tau_A\;,\;x_A=0\;,\;y_A=0\;,\;z_A=z\;,\\
&&t_B=\tau_B\;,\;x_B=0\;,\;y_B=0\;,\;z_B=z+L\; .
\end{eqnarray}
The field two-point correlation function defined in (\ref{eq:12}) and (\ref{eq:13}) is accordingly given by
\begin{eqnarray}
\label{eq:15}
&&g_{ij}(x_A(\tau_A),x_B(\tau_B))={\delta_{ij}\/\pi^2}{{\Delta\tau^2+(1-2n_i)L^2}\/{[(\Delta\tau-i\epsilon)^2-L^2]^3}}\nn\\&&
-{(\delta_{ij}-2n_i n_j)\/\pi^2}{{\Delta\tau^2+(1-2n_i)\mathcal{R}^2}\/{[(\Delta\tau-i\epsilon)^2-\mathcal{R}^2]^3}}\, ,
\end{eqnarray}
where $\Delta\tau=\tau_A-\tau_B$, $L$ is the interatomic distance, and $\mathcal{R}=z_A+z_B=L+2z$ indicates the distance of one atom from the mirror image of the second atom with respect to the plate.
It can be shown that the only nonzero components of $g_{ij}$ are the $xx$, $yy$, and $zz$ components. In particular, due to the symmetry of the physical system considered,  $g_{xx}(x_A(\tau_A), x_B(\tau_B))=g_{yy}(x_A(\tau_A),x_B(\tau_B))\not= g_{zz}(x_A(\tau_A), x_B(\tau_B))$; therefore, the field correlation function evaluated at the position of the two atoms is diagonal but not isotropic. As we will discuss later on, this has nontrivial consequences on the expression of the resonance interaction energy.

The linear susceptibility of the field as a function of frequencies can be easily calculated from (\ref{eq:15}), obtaining
\begin{widetext}
\begin{eqnarray}
\label{eq:16}
&&\chi^{F}_{ij}(x_A(\tau_A),x_B(\tau_B))={1\/8\pi^2}\int^{\infty}_{0}d\omega
\biggl\{\biggl[(\delta_{ij}-3n_in_j)\biggl({\sin(\omega L)\/L^3}-{{\omega}\cos(\omega L)\/L^2}\biggr)-(\delta_{ij}-n_in_j){\omega^2\sin({\omega L})\/L}\biggr]\nn\\&&
\quad\quad\quad\quad+\biggl[f^{\perp}_{ij}(\mathcal{R}){\omega}\cos({\omega\mathcal{R}})-\biggl({f^{\perp}_{ij}(\mathcal{R})\/\mathcal{R}}-{\omega^2} h^{\perp}_{ij}(\mathcal{R})\biggr)\sin({\omega\mathcal{R}})\biggr]\biggr\}(e^{i\omega\Delta\tau}-e^{-i\omega\Delta\tau})\;,
\end{eqnarray}
\end{widetext}
where we have defined
\begin{equation}
\label{eq:17}
f^{\perp}_{ij}(\mathcal{R})=(\delta_{ij}+n_in_j)\mathcal{R}^{-2}\;,\;h^{\perp}_{ij}(\mathcal{R})=(\delta_{ij}-n_in_j)\mathcal{R}^{-1}\;,
\end{equation}
where the superscript $\perp$ refers to the specific configuration we are considering.
Similarly, we can easily obtain the atomic symmetric correlation function
\begin{eqnarray}
\label{eq:18}
C^{AB}_{ij}(\tau,\tau')=\pm {1\/2}(\mu^{A}_{ge})_i(\mu^{B}_{eg})_j(e^{i\omega_0\Delta\tau}+e^{-i\omega_0\Delta\tau})\; ,
\end{eqnarray}
where $\pm$ signs refer to the symmetric or antisymmetric correlated state, respectively.

Substituting (\ref{eq:16}) and (\ref{eq:18}) in (\ref{eq:9}), after some algebra we finally obtain the expression of the resonance interaction between two atoms aligned in the $z$ direction
\begin{eqnarray}
\label{eq:19}
\delta E_{\perp}(L,\mathcal{R})=\delta E_{\perp}^{(0)}(L)+\delta E_{\perp}^{(b)}(\mathcal{R}) \; ,
\end{eqnarray}
where
\begin{widetext}
\begin{eqnarray}
\label{eq:20}
\delta E_{\perp}^{(0)}(L)&=&\pm{1\/4\pi}\frac{1}{L^3}\biggl\{\sum_{i=x,y}(\mu^A_{ge})_{i}(\mu^B_{eg})_{i}\biggl[\cos({\omega_0L})+\omega_0 L\sin({\omega_0L})
-\omega^2_0L^2\cos({\omega_0L})\biggr]\nn\\&
-&2(\mu^A_{ge})_{z}(\mu^B_{eg})_{z}\biggl(\omega_0 L\sin({\omega_0L})
+\cos({\omega_0L})\biggr)\biggr\}\, ,
\end{eqnarray}
and
\begin{eqnarray}
\label{eq:21}
\delta E_{\perp}^{(b)}(\mathcal{R})&=&\pm{1\/4\pi}\frac{1}{\mathcal{R}^3}\biggl\{\sum_{i=x,y}(\mu^A_{ge})_{i}(\mu^B_{eg})_{i}
\biggl[-\cos(\omega_0\mathcal{R})-\omega_0\mathcal{R}\sin(\omega_0\mathcal{R})+\omega^2_0\mathcal{R}^2\cos(\omega_0\mathcal{R})\biggr]\nn\\&
-&2(\mu^A_{ge})_{z}(\mu^B_{eg})_{z}\biggl(\omega_0\mathcal{R}\sin(\omega_0\mathcal{R})+\cos(\omega_0\mathcal{R})\biggr)\biggr\}\, .
\end{eqnarray}
\end{widetext}

Equations (\ref{eq:19} - \ref{eq:21}) give the general expression of the resonance interaction energy as a function of the interatomic distance $L$  and the distance $z$ of atoms from the reflecting plate.
We note the presence of two terms:  $\delta E_{\perp}^{0}(L)$, which is the resonance interaction energy between
two atoms in the free space, and $\delta E_{\perp}^{(b)}(\mathcal{R})$, which is related to the presence of the reflecting plate.  The term $\delta E_{\perp}^{(b)}(\mathcal{R})$ can be physically interpreted as the resonance interaction energy between one atom and the image of the other atom with respect to the plate. The image dipole moment has components $(-\mu_x, -\mu_y, \mu_z).$
This result is valid for arbitrary $L$ and $\mathcal{R}$. In particular, when both atoms are very distant from the mirror, the boundary-dependent term (\ref{eq:21}) goes to zero, and we recover the well-known expression for two atoms in the free space \cite{CT98}.

In the near-zone limit, that is for interatomic distances smaller than the atomic transition wavelength, and for atoms very close to the plate ($k_0L\ll 1$  and $k_0\mathcal{R}\ll1$), Eq. (\ref{eq:19}) reduces to
\begin{eqnarray}
\label{eq:22}
&&\delta E_{\perp}(L,\mathcal{R})\approx\pm{1\/4\pi}\biggl[\sum_{i=x,y}(\mu^A_{ge})_{i}(\mu^B_{eg})_{i}\biggl({1\/L^3}-{1\/\mathcal{R}^3}\biggr)\nn\\&&\quad\quad
-2(\mu^A_{ge})_{z}(\mu^B_{eg})_{z}\biggl({1\/L^3}+{1\/\mathcal{R}^3}\biggr)\biggr]\;,
\end{eqnarray}

Equation (\ref{eq:22}) clearly shows that the resonance interaction can be suppressed or enhanced for specific orientations of the two atomic dipole moments. For example, if the two dipole moments are both oriented in the $x$ (or $y$) direction, which is parallel to the mirror, Eq. (\ref{eq:22}) assumes the form
\begin{eqnarray}
\label{eq:23}
&&\delta E_{\perp}(L,\mathcal{R})\approx\pm{1\/4\pi}(\mu^A_{ge})_{x}(\mu^B_{eg})_{x}\biggl({1\/L^3}-{1\/\mathcal{R}^3}\biggr)\, .
\end{eqnarray}

\begin{figure}[H]
\includegraphics[scale=0.7]{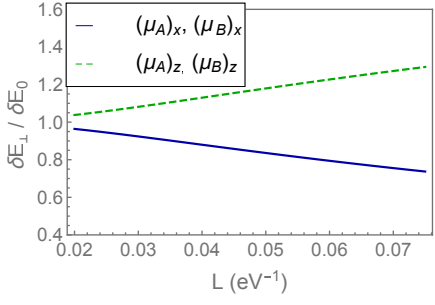}
\caption{Plot of $\delta E/\delta E_0$ as a function of the interatomic distance, $L$, in the near-zone, for atoms aligned perpendicularly to the mirror. The dipole moments are  both assumed oriented in the $z$ direction (green line), or in the $x$ direction (blue line). The interaction energy is enhanced (green-line) or suppressed (blue line) compared to that  in the vacuum, depending on the orientation of the two dipole moments. The parameters are chosen such that $\omega_0=4.17 \text{eV}$, $z_A=2\times10^{-2} \, \text{eV}^{-1}$, $\mu^A_x=\mu^A_z=\mu^B_x=\mu^B_z=1.024\times10^{-3}\text{eV}^{-1}$.}
\label{Figu2}
\end{figure}

A comparison with the expression of the resonance interaction between two atoms in the free space,
\begin{eqnarray}
\label{eq:24}
\delta E_{0}\approx\pm{1\/4\pi}(\mu^A_{ge})_{x}(\mu^B_{eg})_{x}{1\/L^3}\;,
\end{eqnarray}
shows that the presence of the boundary reduces the interaction between the two atoms.
The situation changes if we consider the two dipole moments aligned along the $z$ direction, which is perpendicular to the plate. In this case we get
\begin{eqnarray}
\label{eq:25}
\delta E_{\perp}(L,\mathcal{R})\approx\mp{1\/2\pi}(\mu^A_{ge})_{z}(\mu^B_{eg})_{z}\biggl({1\/L^3}+{1\/\mathcal{R}^3}\biggr)\;,
\end{eqnarray}
while, for atoms in the free space
\begin{eqnarray}
\label{eq:26}
\delta E_{0}\approx\mp{1\/2\pi}(\mu^A_{ge})_{z}(\mu^B_{eg})_{z}{1\/L^3}\;,
\end{eqnarray}
showing that the resonance interaction is enhanced by the presence of the reflecting plate.
These findings, related to the anisotropic behavior of the field susceptibility, are also expected from a physical point of view. When both dipoles are perpendicular to the plate, each dipole and its image form a superradiant
Dicke state and the emission rate is doubled \cite{PT82}. This implies that the resonance interaction energy (which is related to the exchange of a real photon between the atoms) is enhanced. In contrast, if the two dipoles are parallel to the plate, each dipole and its image form a subradiant Dicke state; the spontaneous emission is then inhibited and the resonance interaction energy between the two atoms is weakened.
Figure \ref{Figu2} illustrates the results discussed above, for two atoms of $Rb^{87}$, with $\omega_0=4.17  \,\text{eV}$ and assuming a dipole moment $\mu=1.024 \,\text{eV}^{-1}$ for both atoms
(in our natural units, $1 \; \text{eV}^{-1} \simeq 1.97 \times 10^{-7} \text{m}$.)

\begin{figure}[!htbp]
\centering
\subfigure[]{
\includegraphics[scale=0.60]{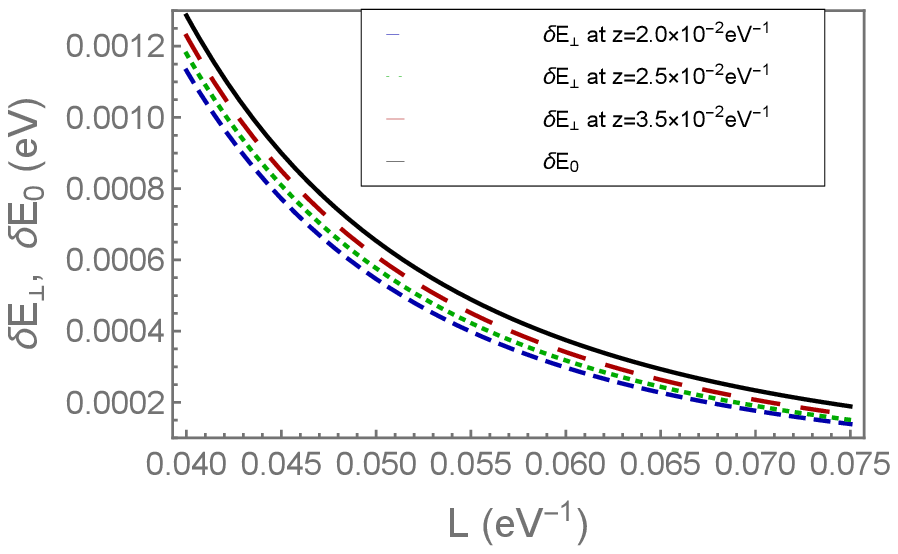}\label{Fig3}}
\subfigure[]{
\includegraphics[scale=0.5]{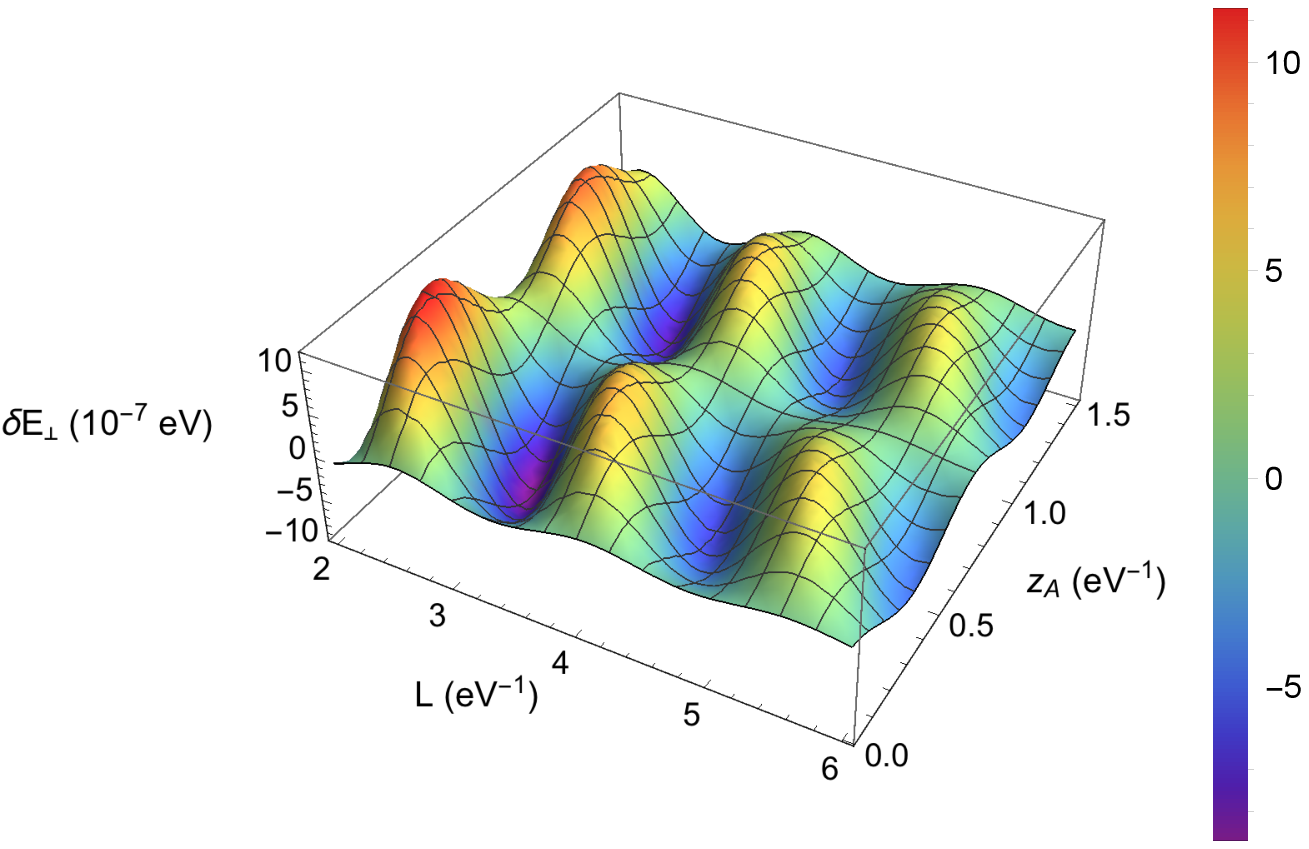}\label{Fig4}}
\subfigure[]{
\includegraphics[scale=0.60]{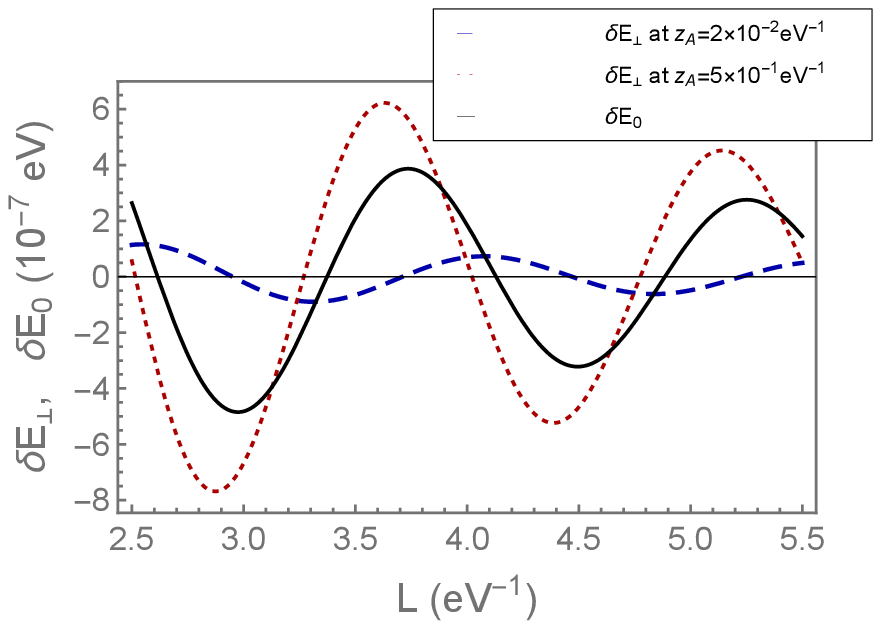}\label{Fig4a}}
\caption{Plots of the resonance interaction energy as a function of the interatomic distance $L$ for atoms aligned perpendicular to the mirror and for different atom-plate distances $z_A$, in (a) the near-zone limit $(L\ll\omega^{-1}_0)$, and (b) and (c) the far-zone limit $(L\gg\omega^{-1}_0)$.  The two dipole moments are both assumed to be oriented in the $x$ direction, with  $\mu^A_x=\mu^B_x=1.024\times10^{-3} \text{eV}^{-1}$; $\omega_0=4.17 \text{eV}$.
(a) The resonance interaction is suppressed with respect to that in vacuum (black solid line) and increases by increasing the atom-plate distance $z_A$. Parameters have been chosen such that
$z_A=2.0\times 10^{-2} \text{eV}^{-1}$ (blue dashed
line), $z_A=2.5\times 10^{-2} \text{eV}^{-1}$ (green dotted line), $z_A=3.5\times 10^{-2} \text{eV}^{-1}$ (red long-dashed line). (b), (c) The resonance interaction has oscillations in space and is enhanced or inhibited compared to the case of atoms {\it in vacuo}, depending on the distances $L$ and  $z_A$ of atom $A$ from the plate. The numerical values of $z_A$ range from $z_A\sim 2.5\times 10^{-2} \text{eV}^{-1} \ll \omega_0^{-1}$ to $z_A\sim 5.5 \text{eV}^{-1}\gg \omega_0^{-1}$.}
\label{figure3}
\end{figure}

Our conclusions are also fully consistent with recent investigations concerning the collective spontaneous decay of two entangled atoms near a perfectly conducting plate \cite{PPRBB17}, where the presence of the boundary can enhance or inhibit the superradiant or subradiant decay of the two-atom system, according to the specific orientation of the two dipole moments and to the atom-plate distances (with respect to the atomic transition wavelength).

Similar features appear in the far-zone limit ($L \gg \omega_0^{-1}$). Also in this case, in fact, the resonance interaction energy can be enhanced or suppressed, depending on the distance of atom $A$ from the plate. Figure $3$ plots the resonance interaction energy between the two atoms, as a function of the interatomic distance, for different distances of atom $A$  from the mirror (the two dipole moments are both assumed to be oriented in the $x$ direction). As shown in the figures, the interaction energy can be suppressed (compared to the free-space case) (Fig.\ref{Fig3}), or enhanced (Fig.\ref{Fig4},\ref{Fig4a}), depending on the distances of atoms $A$ and $B$ from the plate.

\subsection{Two atoms aligned along a direction parallel to the reflecting plate}
\begin{figure}[H]
\includegraphics[scale=0.45]{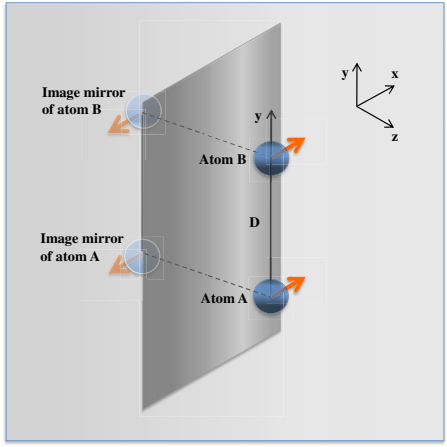}
\caption{Pictorial representation of the physical system of two atoms aligned in the $y$ direction, parallel to the reflecting plate.}\label{Fig5}
\end{figure}

We now assume the two atoms at rest and placed along a direction parallel to the plate, for example in the $y$-direction, separated by a distance $D$ and at a distance $z$ from the plate, as shown in Fig. \ref{Fig5}.

Then
the atomic trajectories are
\begin{eqnarray}
\label{eq:30}
&&t_A=\tau_A\;,\;x_A=0\;,\;y_A=0\;,\;z_A=z\;.\\
&&t_B=\tau_B\;,\;x_B=0\;,\;y_B=D,\;z_B=z\;.
\end{eqnarray}

As before, we first evaluate the two-point correlation function of the electric field operator. We get
\begin{eqnarray}
\label{eq:31}
&\ &g_{ij}(x_A(\tau_A),x_B(\tau_B)) \nonumber \\
&=&g^{(0)}_{ij}(x_A(\tau_A),x_B(\tau_B))+g^{(b)}_{ij}(x_A(\tau_A),x_B(\tau_B)) \; ,
\end{eqnarray}
with
\begin{eqnarray}
\label{eq:32}
g^{(0)}_{ij}(x_A(\tau_A),x_B(\tau_B))={\delta_{ij}\/\pi^2}{{\Delta\tau^2+(1-2p_i)D^2}\/{[(\Delta\tau-i\epsilon)^2-D^2]^3}}
\end{eqnarray}
and
\begin{widetext}
\begin{eqnarray}
\label{eq:33}
g^{(b)}_{ij}(x_A(\tau_A),x_B(\tau_B))=-{1\/\pi^2}{({\delta_{ij}-2n_in_j)[\Delta\tau^2+(1-2p_i)D^2+(1-2n_i)4z^2]-4zD(p_in_j-p_j n_i)}\/{[(\Delta\tau-i\epsilon)^2-R^2]^3}} \; .
\end{eqnarray}
\end{widetext}
Here, $D=y_B-y_A$ is the interatomic distance, $\mathbf{p}=(0,1,0)$ and $R=R(z,D)=\sqrt{D^2+4z^2}$ is the distance between a dipole and the image of the other with respect to the mirror.  The field susceptibility is
\begin{widetext}
\begin{eqnarray}
\label{eq:34}
&\ &\chi^{F}_{ij}(x_a(\tau),x_b(\tau'))={1\/8\pi^2}\int^{\infty}_{0}d\omega
\biggl\{\biggl[(\delta_{ij}-3p_ip_j)\biggl({\sin(\omega D)\/D^3}-{\omega\cos(\omega D)\/D^2}\biggr)-(\delta_{ij}-p_ip_j){\omega^2\sin(\omega D)\/D}\biggr]\nn\\
&\ &\quad\quad\quad\quad
+\biggl[f^{\parallel}_{ij}(z,D)\omega\cos(\omega R)-\biggl({f^{\parallel}_{ij}(z,D)\/R}-\omega^2 h^{\parallel}_{ij}(z,D)\biggr)\sin(\omega R)\biggr]\biggr\}
(e^{i\omega\Delta\tau}-e^{-i\omega\Delta\tau})
\end{eqnarray}
\end{widetext}
where we have defined the tensorial functions
\begin{widetext}
\begin{eqnarray}
\label{eq:35}
f^{\parallel}_{ij}(z,D)&=&[(\delta_{ij}-3p_ip_j-2n_in_j)D^2+4z^2(\delta_{ij}+n_in_j)-6zD(p_in_j-p_j n_i)]R^{-4}\;,\\
h^{\parallel}_{ij}(z,D)&=&[(\delta_{ij}-p_ip_j-2n_in_j)D^2+4z^2(\delta_{ij}-n_in_j)-2zD(p_in_j-p_j n_i)]R^{-3}\;,
\end{eqnarray}
\end{widetext}
where the superscript $\parallel$ indicates the present geometric configuration.
An analysis of Eq. (\ref{eq:34}) immediately shows the presence of two non diagonal terms, $\chi_{yz}^F$ and $\chi_{zy}^F$, not present in the case discussed before (see Eq.\ref{eq:16}). This feature is related to the specific geometric configuration of the system we are considering, in which the translational symmetry in the $y$ and $z$ directions is lost. We now show that this has some relevant consequences on the resonance interaction between the two atoms.

Using Eqs.  (\ref{eq:34}) and (\ref{eq:18}) in Eq. (\ref{eq:9}), after some algebra we obtain the expression of the resonance interaction energy between atoms $A$ and $B$,
\begin{eqnarray}
\label{eq:36}
\delta E_{\parallel}=\delta E_{\parallel}^{(0)}+\delta E_{\parallel}^{(b)} \; ,
\end{eqnarray}
where
\begin{widetext}
\begin{eqnarray}
\label{eq:37}
\delta E_{\parallel}^{(0)}&=&\pm{1\/4\pi D^3}\biggl\{\sum_{i=x,z}(\mu^A_{ge})_{i}(\mu^B_{eg})_{i}\biggl[\cos(\omega_0D)+{\omega_0 D}\sin(\omega_0D)-\omega^2_0D^2\cos(\omega_0D)\biggr]\nn\\
&-&2(\mu^A_{ge})_{y}(\mu^B_{eg})_{y}\biggl(\omega_0 D\sin(\omega_0D)+\cos(\omega_0D)\biggr)\biggr\}\;,
\end{eqnarray}
\end{widetext}
is the usual expression of the interaction energy between two atoms in the free space \cite{CT98} and
\begin{widetext}
\begin{eqnarray}
\label{eq:38}
\delta E_{\parallel}^{(b)}&=&\pm{1\/4\pi}\biggl\{(\mu^A_{ge})_{x}(\mu^B_{eg})_{x}\frac{1}{R^3}\biggl[-\cos(\omega_0R)-\omega_0R\sin(\omega_0R)+\omega^2_0R^2\cos(\omega_0R)\biggr]\nn\\&
+&(\mu^A_{ge})_{z}(\mu^B_{eg})_{z}\frac{1}{R^5}\biggl[(D^2-8z^2)\cos(\omega_0R)+(D^2-8z^2)\omega_0 R\sin(\omega_0R) -D^2\omega^2_0R^2\cos(\omega_0R)\biggr]\nn\\&
+&(\mu^A_{ge})_{y}(\mu^B_{eg})_{y}\frac{1}{R^5}\biggl[2(D^2-2z^2)\cos(\omega_0R)+2(D^2-2z^2)\omega_0 R\sin(\omega_0R)+4z^2\omega^2_0R^2\cos(\omega_0R)\biggr]\nn\\&
+&\biggl((\mu^A_{ge})_{y}(\mu^B_{eg})_{z}-(\mu^A_{ge})_{z}(\mu^B_{eg})_{y}\biggr)\frac{1}{R^5}
\biggl[6zD\omega_0R\sin(\omega_0R)+\biggl(6zD-2zD\omega^2_0R^2\biggr)\cos(\omega_0R)\biggr]\biggr\}\;
\end{eqnarray}
\end{widetext}
is the change due to the presence of the mirror.  A comparison with the previous result for atoms perpendicular to the plate (see Eqs. (\ref{eq:19}) - (\ref{eq:21})) shows the presence of a new term in the expression of the interaction energy, related to the non-diagonal contribution of the field susceptibility. This term is responsible for a qualitative change of the resonance interaction between the two atoms, giving a nonvanishing interaction also when the two dipole moments are not parallel to each other, a result which is at variance with that obtained in the preceding section for atoms aligned perpendicularly to the plate  (see Eq. (\ref{eq:21})).

In the near-zone limit, that is when $D,R\ll\omega_0^{-1}$, the expression (\ref{eq:36}) can be approximated as
\begin{widetext}
\begin{eqnarray}
\label{eq:39}
\delta E_{\parallel}&\approx&{1\/4\pi}\biggl[(\mu^A_{ge})_{x}(\mu^B_{eg})_{x}\biggl({1\/D^3}-{1\/R^3}\biggr)+(\mu^A_{ge})_{y}(\mu^B_{eg})_{y}\biggl(-{2\/D^3}+{2D^2-4z^2\/D^5}\biggr)\nn\\&
+&(\mu^A_{ge})_{z}(\mu^B_{eg})_{z}\biggl({1\/D^3}+{D^2-8z^2\/R^5}\biggr)+((\mu^A_{ge})_{y}(\mu^B_{eg})_{z}-(\mu^A_{ge})_{z}(\mu^B_{eg})_{y}){6zD\/R^5}\biggr]\; .
\end{eqnarray}
\end{widetext}

\begin{figure}[!htpb]
\centering
\subfigure[]{
\includegraphics[scale=0.6]{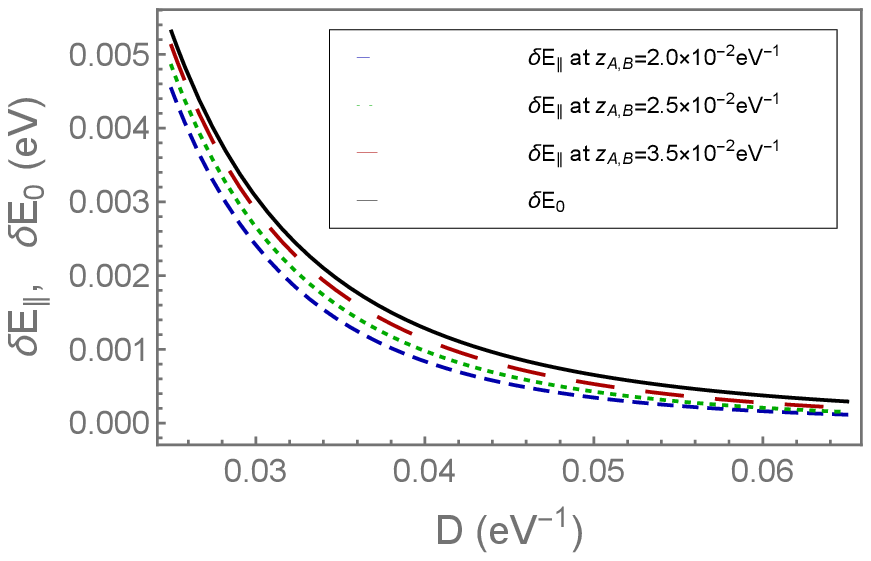}\label{Fig6}}
\subfigure[]{
\includegraphics[scale=0.55]{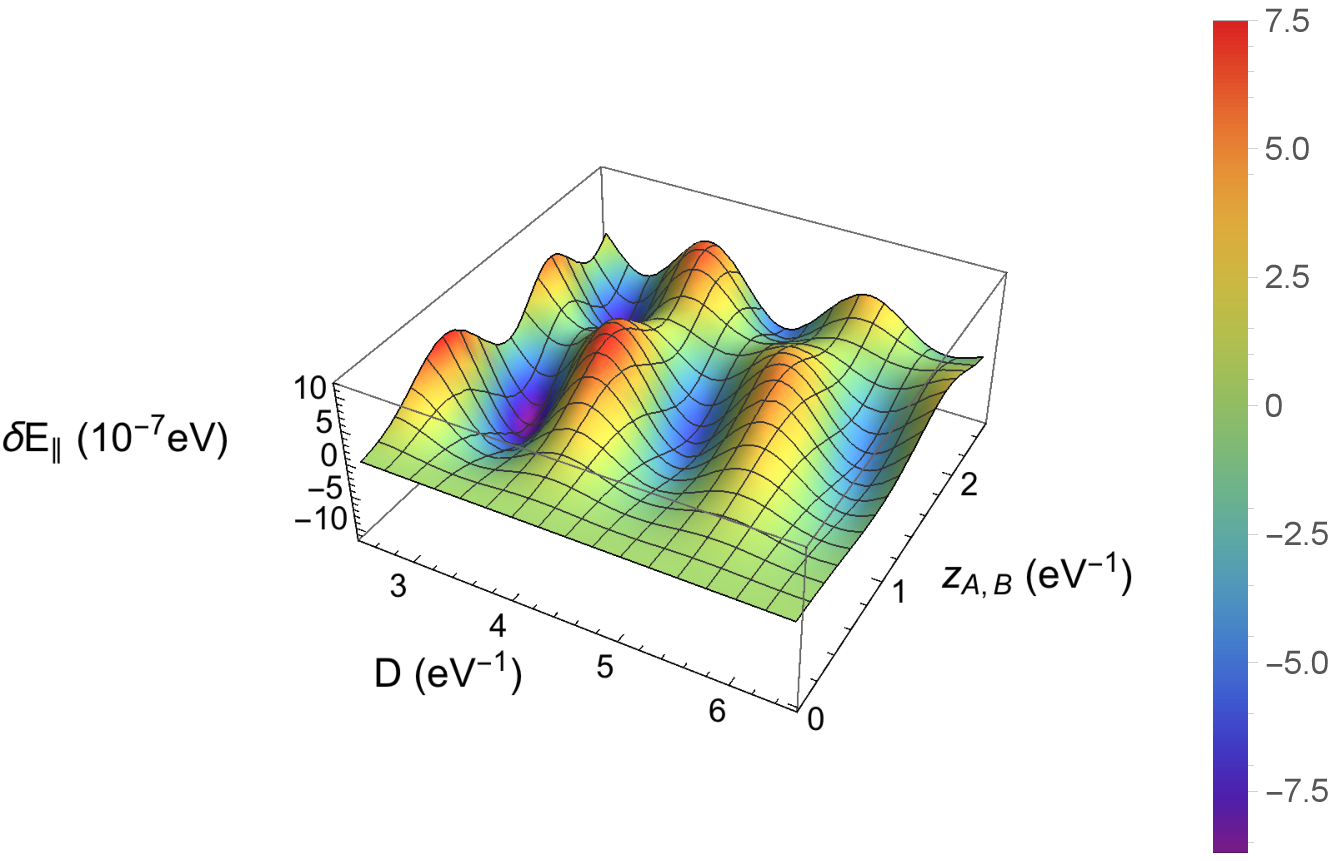}\label{Fig7}}
\subfigure[]{
\includegraphics[scale=0.6]{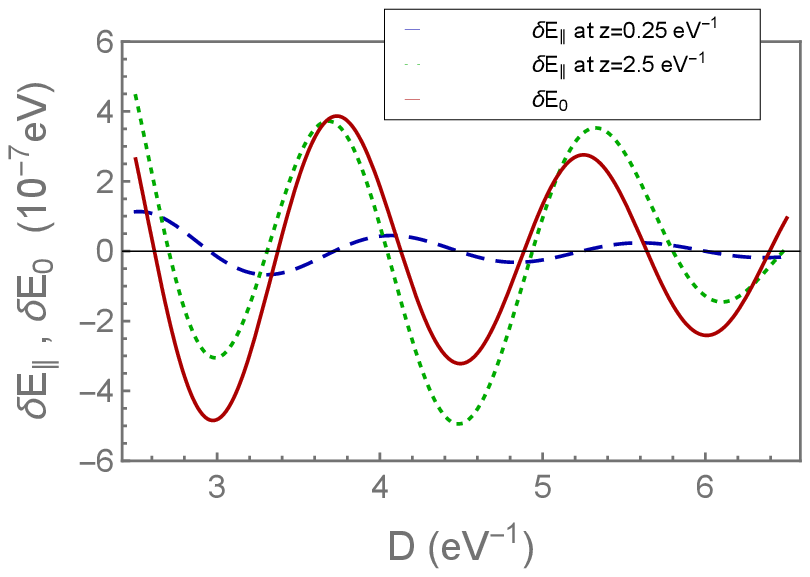}\label{Fig9c}}
\caption{Plots of the resonance interaction energy as a function of the interatomic distance $D$, for atoms aligned parallel to the plate in (a) the near-zone limit ($D\ll\omega_{0}^{-1}$) and for atoms very close to the plate and (b) and (c) the far-zone limit $D\gg\omega_{0}^{-1}$, at different atom-plate distances. The dipole moments are both assumed to be oriented in the $x$ direction. (a) The interaction is inhibited by the presence of the mirror, and approaches that of atoms in vacuum (black solid line) when the distance $z$ of atoms $A$ and $B$ from the mirror is increased. The parameters are chosen such that $z=2.0\times 10^{-2} \text{eV}^{-1}$ (blue dashed line), $2.5\times 10^{-2} \text{eV}^{-1}$ (green dot-dashed line), $z=3.5\times 10^{-2} \text{eV}^{-1}$ (red dotted line), $\omega_0=4.17 \text{eV}$, $\mu^A_x=\mu^B_x=1.024\times10^{-3} \text{eV}^{-1}$. The black solid line represents the resonance interaction between two atoms in vacuum.
(b) and (c) The resonance interaction shows oscillations in space and can be suppressed or enhanced with respect to that between atoms in the unbounded space, by varying the distances involved. The parameters are chosen such that $\omega_0=4.17 \text{eV}$,   $\mu^A_x=\mu^B_x=1.024\times10^{-3} \text{eV}^{-1}$, $2.5\times 10^{-2} \leq D\leq 6.5$ $\text{eV}^{-2}$.}
\label{figure2}
\end{figure}

Figure \ref{Fig6} shows a plot of Eq.  (\ref{eq:39}) as a function of the interatomic distance $D$ for different atom-plate distances, $z$. It shows that the interaction decays as a power law,  and it is strongly suppressed by the presence of the mirror, compared to the case of atoms in vacuum; as expected, the effect of the mirror decreases when the atom-plate distance $z$ increases.

Similar effects manifest also in the far-zone limit, $D\gg\omega_{0}^{-1}$, where the interatomic interaction can be suppressed or enhanced by the presence of the reflecting plate, depending on the distances $D$ and $z$, as shown in Figs. \ref{Fig7} and \ref{Fig9c}.

\begin{figure}[!htbp]
\centering
\subfigure[]{
\includegraphics[scale=0.6]{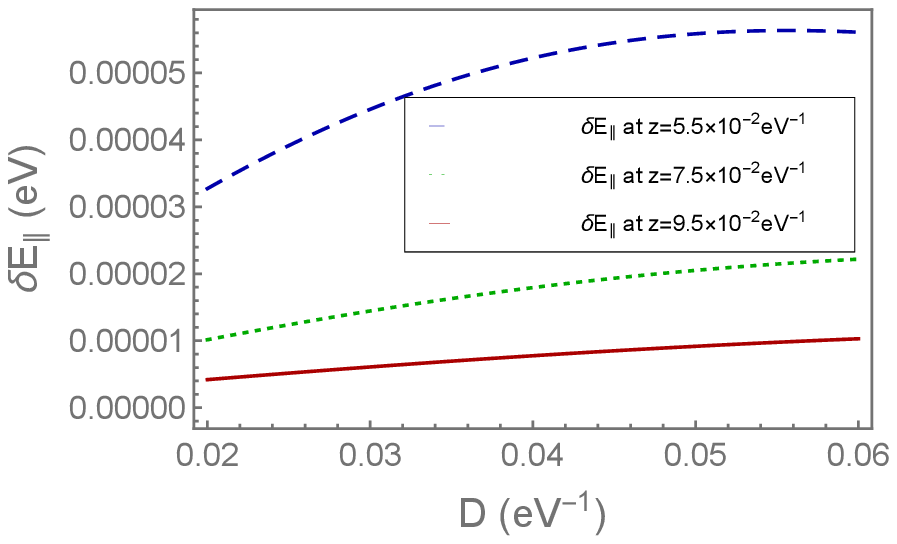}\label{Fig8}}
\subfigure[]{
\includegraphics[scale=0.6]{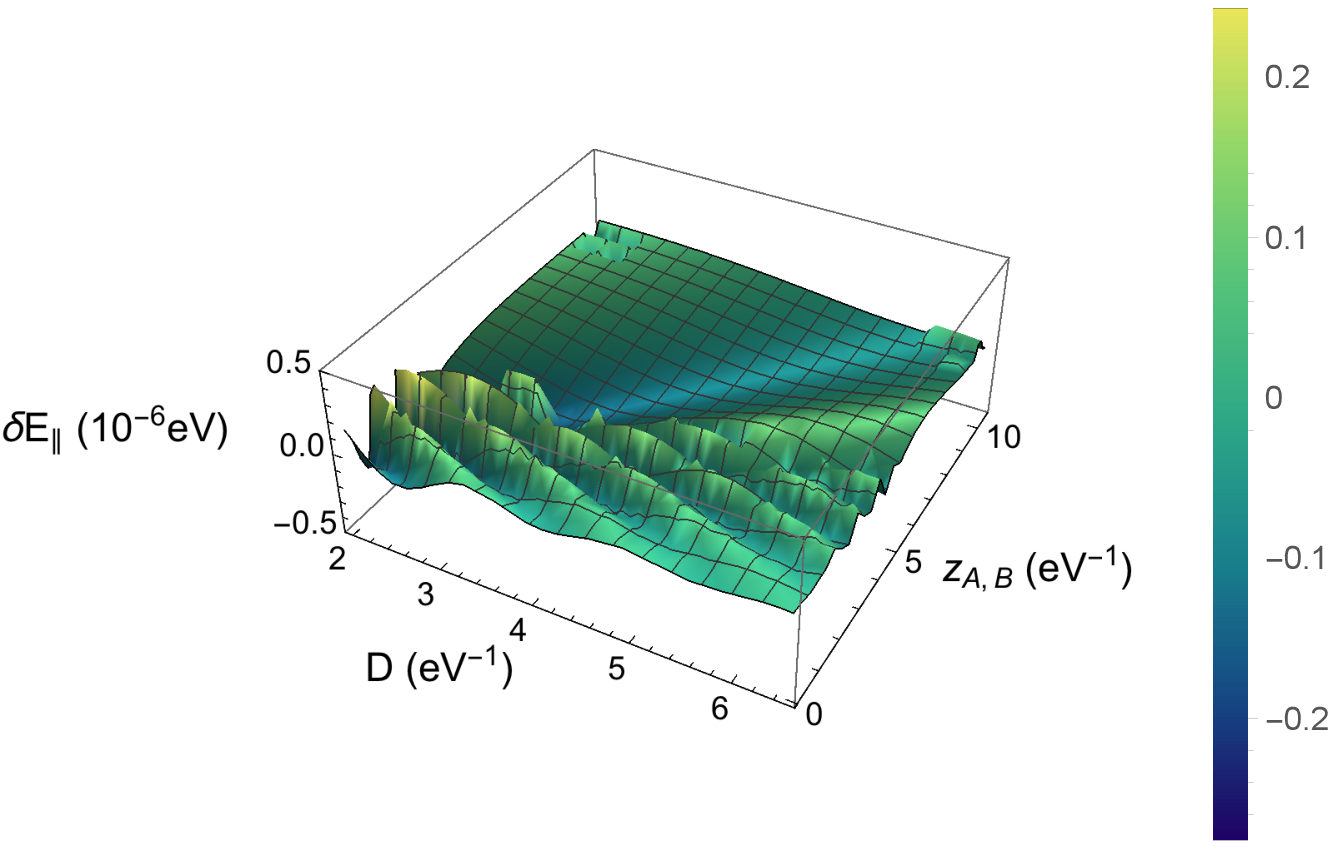}\label{Fig9}}
\caption{Plots of the resonance interaction energy as a function of the interatomic distance $D$ for atoms with dipole moments along $y$ and $z$ in (a) the near-zone limit ($D\ll\omega_{0}^{-1}$) and (b) the far-zone limit ($D\gg\omega_{0}^{-1}$), with $\mu_A$ assumed to be in the $y$ direction and $\mu_B$ in the $z$ direction. The parameters are chosen such that $\omega_0=4.17 \text{eV}$, $\mu^A_y=\mu^B_z=1.024\times 10^{-3}\text{eV}^{-1}$. (a) The interaction  decreases by increasing the distance $z$ of atoms $A$ and $B$ from the mirror.
(b) The interaction energy oscillates with the distance and it is of the same order of magnitude of that for atoms in the vacuum.}
\end{figure}
\begin{figure}[!htpb]
\centering
\includegraphics[scale=0.7]{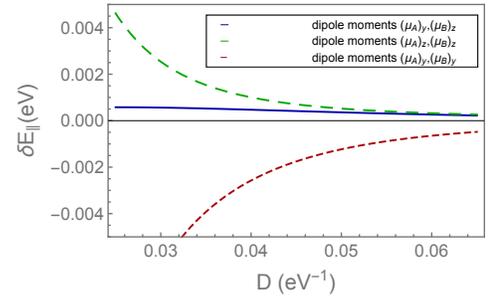}
\caption{Comparison between the resonance energies as a function of the interatomic distance D, for different orientations of the dipole moments. The atoms are assumed to be very close to the mirror. The plot shows that the interaction decays more slowly with distance compared to the case of parallel dipole moments. The values of the parameters used are $z=2.5\times 10^{-2} \text{eV}^{-1}$, $\omega_0=4.17 \text{eV}$,   $\mu^A_x=\mu^B_x=1.024\times10^{-3} \text{eV}^{-1}$.}
\label{Fig10}
\end{figure}

The results discussed above are similar to those illustrated in the preceding section for the perpendicular configuration. Interesting effects appear if we consider the case of two dipole moments oriented perpendicular to each other, specifically one in the $y$-direction and the other in the $z$-direction. In fact, in this case an extra term appears, not present in the well-known case of atoms in the unbounded space, related to the specific configuration of the system. This term can give a change of the resonance interaction energy in specific configurations. Figures \ref{Fig8} and \ref{Fig9} show the behavior of the resonance interaction energy in the near- and far-zone limits, for different atom-plate distances. It is worth observing that, in both cases, the strength of the interaction is of the same order of magnitude as the resonance interaction energy between two atoms in vacuum, thus suggesting the possibility to observe the peculiar effects related to the presence of the reflecting  plate and the specific geometric configuration analyzed. Figure \ref{Fig10} shows a comparison between the resonance energies in the near-zone limit, for different orientations of the atomic dipole moments.

In conclusion, our findings clearly show that the interaction energy can be strongly affected by the presence of an external environment, such as a reflecting mirror.
Even if our model of two-level atoms near a perfectly reflecting boundary is somehow idealized, the main approximations used are not far from a realistic experimental setup. In fact, the two-level approximation is justified by the fact that the most relevant transitions are resonant and thus other atomic levels do not significantly contribute. Also, if the transition frequency of the two identical atoms is below the plasma frequency of a realistic metallic boundary, as is the case of transitions in the optical region and metals such as gold (whose plasma frequency is in the ultraviolet region of the spectrum), also the approximation of a perfectly reflecting boundary is reasonable. The two atoms could be maintained near the boundary using available atomic trapping techniques for cold atoms with micrometer scale precision \cite{Reinhard08,Pillet09,Antezza14}.
Since the resonance interaction is strictly related to the energy transfer process between two atoms, these results could be relevant to activate or inhibits such process. In particular, the presence of a {\em non-diagonal} term of the interaction energy, related to the presence of the plate, suggests the possibility to activate the excitation transfer between two atoms, even when it is inhibited in the free space.

\section{Conclusions}
\label{Sec.4}
In this paper we have investigated the resonance interaction energy between two identical two-level atoms, one in the excited state and the other in the ground state, prepared in a correlated Bell-type state, near a perfectly reflecting boundary.
We considered the contribution of vacuum fluctuations and radiation reaction to the resonance interaction energy, and shown that only the radiation reaction term contributes.
We gave a simple physical explanation for this result.
We then considered different geometric configurations for the two-atom system and for different orientations of the atomic dipole moments with respect to the plate.
In particular, we have considered  the cases of two atoms aligned in a direction perpendicular or parallel to the boundary We showed that in the cases considered the resonance interaction is significantly modified  by the presence of the boundary. We discussed in detail, for specific geometric configurations of the two-atom system and orientations of the atomic dipole moments, that it exhibits a change of its strength, of its  distance dependence due to the presence of the boundary, as well as the possibility of enhancing or inhibiting the interatomic interaction.
In particular, we showed, in the case of atoms parallel to the reflecting plate, that the presence of the boundary can give a nonvanishing interaction energy even for configurations of the atomic dipoles for which the interaction between the atoms in free space is zero.
We discussed the physical origin of  these  results, and stressed their relevance from an experimental point of view and also the possibility to activate or inhibit the resonance energy transfer between atoms.

\begin{acknowledgments}
W.Z. is grateful for financial support from the National Natural Science Foundation of China (NSFC) through Grants No. 11405091, No. 11690034, No. 11375092,  and No. 11435006. Key Laboratory of Low Dimensional Quantum Structures and Quantum Control of Ministry of Education through Grant No. QSQC1525, China Scholarship Council (CSC), the Research program of Ningbo University through Grants No. XYL18027, and K. C. Wong Magna Fund in Ningbo University.
R.P. and L.R. gratefully acknowledge financial support from the Julian Schwinger Foundation.
\end{acknowledgments}

\end{document}